\documentclass[conference, a4paper]{IEEEtran}

\usepackage{algorithm,algorithmic}
\usepackage{amsmath,amssymb,mathrsfs}
\usepackage{array}
\usepackage{balance}
\usepackage{booktabs}
\usepackage{cite}
\usepackage{epsfig}
\usepackage{fancyhdr}
\usepackage{float}
\usepackage[nomain,acronym,toc,nopostdot]{glossaries}
\usepackage{graphicx,color}
\usepackage{mathtools}
\usepackage{multirow}
\usepackage{psfrag}
\usepackage{stfloats}
\usepackage{subfigure}
\usepackage{tikz,siunitx}
\usepackage{url}
\usepackage{svg}
\usepackage[utf8]{inputenc}
\usepackage[T1]{fontenc}

\newtheorem{cor}{Corollary}

\newtheorem{prop}{Proposition}

\newtheorem{rmk}{Remark}

\makeglossaries
\newacronym{1d}{1D}{one-dimensional}
\newacronym{2d}{2D}{two-dimensional }
\newacronym{3d}{3D}{three-dimensional}
\newacronym{3gpp}{3GPP}{3rd Generation Partnership Project}
\newacronym{5g}{5G}{fifth-generation}
\newacronym{6g}{6G}{sixth-generation}

\newacronym{ap}{AP}{access point}
\newacronym{amp}{AMP}{approximate message passing}
\newacronym{aoa}{AoA}{angle of arrival}
\newacronym{aod}{AoD}{angle of departure}
\newacronym{ask}{ASK}{amplitude shift keying}
\newacronym{awgn}{AWGN}{additive white Gaussian noise}

\newacronym{ber}{BER}{bit error rate}
\newacronym{bler}{BLER}{bit error rate}
\newacronym{bp}{BP}{belief propagation}
\newacronym{bpsk}{BPSK}{binary phase shift keying}
\newacronym{bs}{BS}{base station}

\newacronym{cbsm}{CBSM}{correlation-based stochastic model}
\newacronym{csirs}{CSI-RS}{channel state information reference signal}
\newacronym{cdf}{CDF}{cumulative distribution function}
\newacronym{cdma}{CDMA}{code division multiple access}
\newacronym{cg}{CG}{conjugate gradient descent}
\newacronym{clt}{CLT}{central limit theorem}
\newacronym{csi}{CSI}{channel state information }
\newacronym{cvp}{CVP}{closest vector problem}

\newacronym{dof}{DoF}{degrees of freedom}

\newacronym{edw}{EDW}{exponentially decaying window}
\newacronym{elaa}{ELAA}{extremely large aperture array}
\newacronym{etsi}{ETSI}{European Telecommunications Standards Institute}

\newacronym{ff}{FF}{far-field}
\newacronym{fcsd}{FCSD}{fixed-complexity sphere decoder}
\newacronym{fec}{FEC}{forward error correction}
\newacronym{fspl}{FSPL}{free space path loss}

\newacronym{gbsm}{GBSM}{geometry-based stochastic model}
\newacronym{gmsk}{GMSK}{Gaussian minimum shift keying}
\newacronym{gs}{GS}{Gauss-Seidel}
\newacronym{gsm}{GSM}{global system for mobile communication}

\newacronym{iid}{i.i.d.}{independently and identical distributed}
\newacronym{ils}{ILS}{integer least-squares}
\newacronym{ind}{i.n.d.}{independently and non-identical distributed}
\newacronym{imt}{IMT}{International Mobile Telecommunications}
\newacronym{iou}{IoU}{intersection over union}
\newacronym{isac}{ISAC}{integrated sensing and communication}
\newacronym{isi}{ISI}{intersymbol interference}
\newacronym{itur}{ITU-R}{International Telecommunication Union Radiocommunication Sector}
\newacronym{iui}{IUI}{inter-user interference}

\newacronym{jsac}{JSAC}{joint sensing and communication}

\newacronym{las}{LAS}{likelihood ascent search}
\newacronym{lbfgs}{L-BFGS}{limited-memory Broyden–Fletcher–Goldfarb–Shanno}
\newacronym{lll}{LLL}{Lenstra-Lenstra-Lov\'{a}sz}
\newacronym{llr}{LLR}{log-likelihood ratio}
\newacronym{los}{LoS}{line-of-sight}
\newacronym{lr}{LR}{lattice reduction}
\newacronym{lmmse}{LMMSE}{minimum mean square error}
\newacronym{lsd}{LSD}{list sphere decoder}
\newacronym{lte}{LTE}{long-term evolution}
\newacronym{ls}{LS}{least-squares}

\newacronym{map}{MAP}{maximum a posteriori probability}
\newacronym{mf}{MF}{matched filter}
\newacronym{mimo}{MIMO}{multiple-input multiple-output}
\newacronym{mld}{MLD}{maximum likelihood detection}
\newacronym{mmimo}{mMIMO}{massive multiple-input multiple-output}
\newacronym{mmwave}{mmWave}{millimeter wave}
\newacronym{mrc}{MRC}{maximum ratio combining}
\newacronym{mse}{MSE}{mean square error}
\newacronym{mt}{MT}{mobile terminal}
\newacronym{nf}{NF}{near-field}
\newacronym{nlos}{NLoS}{non-LoS}

\newacronym{od}{OD}{orthogonality defect}
\newacronym{ofdm}{OFDM}{orthogonal frequency-division multiplexing}

\newacronym{pdf}{PDF}{probability density function}
\newacronym{pda}{PDA}{probabilistic data association}
\newacronym{pep}{PEP}{pairwise error probability}
\newacronym{pl}{PL}{path-loss}
\newacronym{pmf}{PMF}{probability mass function}
\newacronym{pwm}{PWM}{plane-wave model}

\newacronym{qam}{QAM}{quadrature amplitude modulation}
\newacronym{qpsk}{QPSK}{quadrature phase shift keying}

\newacronym{rts}{RTS}{reactive tabu search}
\newacronym{ris}{RIS}{reconfigurable intelligent surface}
\newacronym{rss}{RSS}{received signal strength}
\newacronym{rzf}{RZF}{regularized-ZF}

\newacronym{sa}{SA}{Seysen's algorithm}
\newacronym{sd}{SD}{sphere decoding}
\newacronym{sdr}{SDR}{semidefinite relaxation}
\newacronym{ser}{SER}{symbol error rate}
\newacronym{sf}{SF}{shadow fading}
\newacronym{sic}{SIC}{succesive interference cancellation}
\newacronym{sinr}{SINR}{signal-to-interference-plus-noise ratio}
\newacronym{siso}{SISO}{single input single output}
\newacronym{snr}{SNR}{signal-to-noise ratio}
\newacronym{sns}{SNS}{spatial non-stationarity}
\newacronym{sota}{SoTA}{state-of-the-art}
\newacronym{ssor}{SSOR}{symmetric successive over-relaxation}
\newacronym{svd}{SVD}{singular value decomposition }
\newacronym{swm}{SWM}{spherical-wave model}

\newacronym{tof}{ToF}{time of flight}
\newacronym{thz}{THz}{Terahertz}
\newacronym{tr}{TR}{Technical Report}
\newacronym{ts}{TS}{tabu search}

\newacronym{uca}{UCA}{uniform cylindrical array}
\newacronym{ue}{UE}{user equipment}
\newacronym{ula}{ULA}{uniform linear array}
\newacronym{ura}{URA}{uniform rectangular array}
\newacronym{uma}{UMa}{urban macro}
\newacronym{umi}{UMi}{urban micro}
\newacronym{upa}{UPA}{uniform planar array}

\newacronym{vblast}{V-BLAST}{vertical Bell Labs layered space-time}

\newacronym{xlmimo}{XL-MIMO}{extra-large multiple-input multiple-output}

\newacronym{zf}{ZF}{zero-forcing}

\definecolor{sbluser}{RGB}{0,51,120}
\definecolor{sred}{RGB}{200,51,130}

\newcommand{\appref}[1]{\textsc{Appendix} \ref{#1}}
\newcommand{\corref}[1]{{\it Corollary \ref{#1}}}

\renewcommand{\eqref}[1]{(\ref{#1})}
\newcommand{\figref}[1]{Fig. \ref{#1}}

\newcommand{\propref}[1]{{\it Proposition \ref{#1}}}
\newcommand{\remref}[1]{{\it Remark \ref{#1}}}

\ifCLASSINFOpdf
\else
\fi
\begin{document}
\title{ELAA-ISAC: Environmental Mapping Utilizing the LoS State of Communication Channel}
\author{Jiuyu Liu, Chunmei Xu, Yi Ma$^\dag$, Rahim Tafazolli, and Ahmed Elzanaty\\
	{\small 5GIC and 6GIC, Institute for Communication Systems, University of Surrey, Guildford, UK, GU2 7XH}\\
	{\small Emails: (jiuyu.liu, chunmei.xu, y.ma, r.tafazolli, a.elzanaty)@surrey.ac.uk}}
\markboth{}%
{}

\maketitle

\begin{abstract}
In this paper, a novel environmental mapping method is proposed to outline the indoor layout utilizing the \gls{los} state information of \gls{elaa} channels.
It leverages the spatial resolution provided by ELAA and the \gls{mt}'s mobility to infer the presence and location of obstacles in the environment.
The LoS state estimation is formulated as a binary hypothesis testing problem, and the optimal decision rule is derived based on the likelihood ratio test.
Subsequently, the theoretical error probability of LoS estimation is derived, showing close alignment with simulation results.
Then, an environmental mapping method is proposed, which progressively outlines the layout by combining LoS state information from multiple \gls{mt} locations.
It is demonstrated that the proposed method can accurately outline the environment layout, with the mapping accuracy improving as the number of service-antennas and \gls{mt} locations increases.
This paper also investigates the impact of channel estimation error and \gls{nlos} components on the quality of environmental mapping.
The proposed method exhibits particularly promising performance in \gls{los} dominated wireless environments characterized by high Rician $K$-factor. 
Specifically, it achieves an average \gls{iou} exceeding $80\%$ when utilizing $256$ service antennas and $18$ MT locations.
\end{abstract}                                                                           \glsresetall
                       
\glsresetall

\section{Introduction}\label{sec1}
\Gls{isac} leverages shared infrastructure and signal processing methods for simultaneous data transmission and sensing \cite{Liu2022b}.
\Gls{elaa}, initially investigated for enhancing communication throughput, demonstrates capability for \gls{isac} applications by offering high spatial-resolution \cite{Cui2023}.
Recent research indicates that ELAA channels may consist of a mixture of \gls{los} and \gls{nlos} links \cite{Liu2021, Lu2023, Liu2024}.
Traditionally, obstacles obstructing \gls{los} paths between transmitters and receivers were deemed detrimental to communication \cite{Qing2024}.
However, the spatial distribution of \gls{los} state is inherently linked to the environmental configuration.
The wireless channel requires estimation for communication purposes.
Inferring the LoS state from this estimated channel does not compromise communication performance.
This differs from most ISAC research, which requires performance tradeoffs between communication and sensing applications \cite{GonzalezPrelcic2024, An2023}.

\begin{figure}[t]
	\centering
	\includegraphics[width=0.80\linewidth]{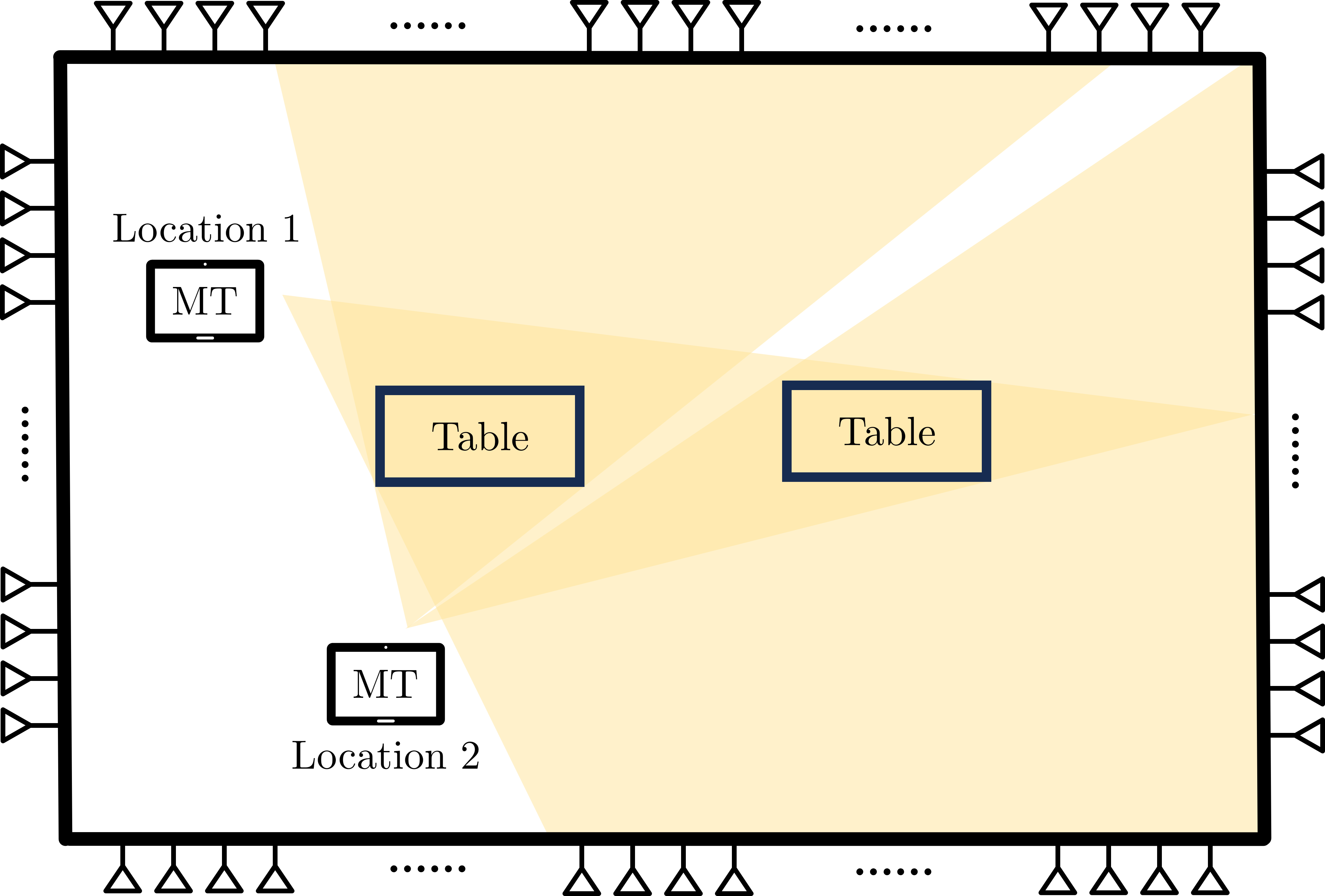}
	\caption{\label{fig01} An indoor environment deployed with an ELAA system. The object being sensed may obstruct the LoS path between the \gls{mt} and ELAA antennas.}
	\vspace{-1em}
\end{figure}

The dense spatial sampling of LoS state information provided by ELAA presents a unique opportunity for ISAC \cite{Zhi2024,Elzanaty2024}.
Research on this communication-centric \gls{isac} for environmental mapping remains in its early stages.
To illustrate how LoS state information from ELAA channels can be used for sensing applications, an example is provided in \figref{fig01}.
The figure depicts an indoor environment with a rectangular ELAA and a \gls{mt}.
Potential \gls{los} paths appear as light yellow areas extending from specific ELAA antennas to various \gls{mt} positions.
However, these paths are blocked obstacles, i.e., the two tables, creating \gls{nlos} regions.
This visualization demonstrates how the \gls{los} state between ELAA and \gls{mt} is influenced by the room configuration and \gls{mt} location.
The locations of objects can be inferred through analysis of LoS state information.
This establishes the foundation for employing the LoS state of ELAA channels in environmental mapping.

In this paper, an environmental mapping method is proposed to outline the indoor layout, utilizing the LoS state of ELAA channels.
The algorithm exploits ELAA spatial resolution and \gls{mt} mobility to infer obstacle presence and location.
LoS state estimation is formulated as a binary hypothesis testing problem, with optimal and practical solutions derived.
Subsequently, the theoretical error probability of LoS state estimation is derived, demonstrating close alignment with simulation results.
The proposed environmental mapping method operates on the principle that LoS links exist between the \gls{mt} and ELAA antennas signify obstacle-free paths.
Our simulation results demonstrate that in \gls{los} dominated environment characterized by high $K$-factor (more than $15$ dB), the proposed algorithm can achieve a high \gls{iou} about $80\%$.
Future research in this area could focus on enhancing LoS estimation techniques, and considering multiple \glspl{mt} with moving targets.

\section{ELAA Channel Model, Channel Estimation, and Problem Statement}\label{sec2}
\subsection{ELAA Channel Model}
MT is located in the near field of the ELAA and the wireless channel should be characterized by a spherical wave model.
Specifically, the path-loss ($p_m$) varies along the antenna array, and channel allows a mixture of \gls{los} and \gls{nlos} links.
For an ELAA with $M$ service antennas, the narrowband wireless channel is expressed as follows \cite{Liu2024a, Amiri2022}
\begin{equation} \label{eqn10380922}
	\mathbf{h}= \sqrt{\mathbf{p}} \odot \bigg(\mathbf{b} \odot \sqrt{\dfrac{K}{K + 1}} \mathbf{h}^{\textsc{los}} + \sqrt{\dfrac{1}{K + 1}} \mathbf{h}^{\textsc{nlos}}\bigg),
\end{equation}
where $\mathbf{h} \in \mathbb{C}^{M \times 1}$ represents the random channel vector, $\odot$ the element-wise product, $b_{m} \in \{0, 1\}$ the LoS state between the MT and $m^{th}$ ELAA antenna, $K$ the Rician $K$-factor, $\mathbf{h}^{\textsc{los}} \in \mathbb{C}^{M \times 1}$ the phase of LoS path, $\mathbf{h}^{\textsc{nlos}} \sim \mathcal{CN}(\mathbf{0}, \mathbf{I}_{M})$ the NLoS components, and $\mathbf{I}_{M}$ denotes an $(M) \times (M)$ identity matrix.

\subsection{Channel Estimation}
For communication purpose, the \gls{mt} transmits pilot signals to the \gls{elaa} for channel estimation.
The channel estimation error typically follows a complex Gaussian distribution as \cite{Liu2024b}
\begin{equation} \label{eqn02440914}
	\widehat{\mathbf{h}} = \mathbf{h} + \mathbf{v},
\end{equation}
where $\widehat{\mathbf{h}}\in \mathbb{C}^{M \times 1}$ denotes the estimated channel, and $\mathbf{v} \sim \mathcal{CN}(0, \sigma_{v}^{2}\mathbf{I}_{M})$ is the estimation error with variance $\sigma_{v}^{2}$.
In practice, the error variance $\sigma_{v}^{2}$ can be reduced by transmitting pilot signals with higher power.

\subsection{Problem Statement} \label{sec1c}
The objective of this paper focuses on environmental mapping using LoS state information from ELAA channels.
This objective encompasses two distinct problems: \textit{1)} Given the estimated channel vector, determining the LoS state of each channel link accurately and efficiently; and \textit{2)} Developing a comprehensive environmental map by leveraging the estimated LoS state information.
These challenges are interconnected, as the accuracy of LoS state estimation directly impacts the quality of environmental mapping.

\section{The Proposed Environmental Mapping Method}
To address the two problems above, this section is divided into two subsections: the first focuses on LoS state estimation and the second focuses on environmental mapping methods.

\subsection{LoS State Estimation}
The objective is to estimate the LoS state of each channel link.
To facilitate mathematical presentation, the $m^{th}$ estimated channel element ($\hat{h}_{m}$) can be expressed as follows
\begin{equation} \label{eqn09270915}
	\hat{h}_{m} = b_{m} \phi_{m} + \omega_{m},
\end{equation}
where $\phi_{m}$ and $\omega_{m}$ are defined as follows
\begin{subequations}
	\begin{align}
		\phi_{m} &\triangleq \sqrt{\dfrac{Kp_{m}}{K + 1}} h_{m}^{\textsc{los}}; \label{eqn08220926} \\
		\omega_{m} &\triangleq \sqrt{\dfrac{p_m}{K + 1}}h_{m}^{\textsc{nlos}} + v_{m}, \label{eqn04560925}
	\end{align}
\end{subequations}
where $\phi_{m}$ denotes the direct LoS path, $h_{m}^{\textsc{los}} = \exp(-j\frac{2\pi}{\lambda}d_{m})$ the phase of LoS path, $d_{m}$ the Euclidean distance between the $m^{th}$ ELAA antenna and \gls{mt}, and $\omega_{m}$ encompasses both the \gls{nlos} components and channel estimation error.

Given \eqref{eqn09270915}, the LoS state estimation can be formulated as the following binary hypothesis testing
\begin{subequations}\label{eqn02060926}
	\begin{align}
		\mathcal{H}_{0}: \quad & b_{m} = 0; \label{eqn02060926a} \\
		\mathcal{H}_{1}: \quad & b_{m} = 1, \label{eqn02060926b}
	\end{align}
\end{subequations}
where $\mathcal{H}_{0}$ denotes the NLoS hypothesis, and $\mathcal{H}_{1}$ denotes the LoS hypothesis.
The LoS state estimation method in this paper is based on the following two assumptions:

\textit{A1):} The locations of MT and ELAA antennas are assumed to be known, providing known values for $d_{m}, ~_{\forall m}$.
Based on the 3GPP document, $p_m$ is expressed as follows \cite{3GPP2022}
\begin{equation}
	p_{m} = 32.4 + 17.3\log_{10}(d_{m}) + 20 \log_{10}(f_{c}),
\end{equation}
where $f_{c}$ denotes the carrier frequency known to the system.
Consequently, the path-loss vector $\mathbf{p}$ can be determined.

\textit{A2):} The Rician $K$-factor and $\sigma_{v}^{2}$ are assumed to be known parameters.
Since $\omega_{m}$ represents the sum of two zero mean complex Gaussian random variables, it follows a complex Gaussian distribution with zero mean and variance
\begin{equation} \label{eqn08300925}
	\sigma_{\omega_{m}}^{2} = \dfrac{p_{m}}{K + 1}  + \sigma_{v}^{2},
\end{equation}
where $p_{m}$ can be determined based on assumption \textit{A1)}.
Hence, $\sigma_{\omega_{m}}^{2}$ constitutes a known parameter.
Also, the LoS path $\phi_{m}$ can be determined according to its definition in \eqref{eqn08220926}.

\begin{prop} \label{prop01}
	Given $\hat{h}_{m}$, and assumptions \textit{A1)} and \textit{A2)}, the decision rule for LoS state estimation is expressed as follows
	\begin{equation} \label{eqn05260925}
		\hat{b}_{m} =\Big(\big|\hat{h}_{m}\big|^{2} - \big|\hat{h}_{m} - \phi_{m}\big|^{2}\Big) \underset{\mathcal{H}_{0}}{\overset{\mathcal{H}_{1}}{\gtrless}} \Theta_{m},
	\end{equation}
	where $\hat{b}_{m}$ denotes the estimated LoS state, and $\Theta_{m}$ is the threshold for the test.
	The optimal $\Theta_{m}$ that minimizes the error probability of LoS state estimation is given by
	\begin{equation}\label{eqn11031105}
		\Theta_{m}^{\star} = \sigma_{\omega_{m}}^{2} \ln\bigg(\frac{\mathcal{P}{(\mathcal{H}_{1})}}{\mathcal{P}{(\mathcal{H}_{0}\big)}}\bigg)
	\end{equation}
	where $P(\mathcal{H}_{0})$ and $P(\mathcal{H}_{1})$ are the prior probabilities of $\mathcal{H}_{0}$ and $\mathcal{H}_{1}$, respectively, with $P(\mathcal{H}_{0}) + P(\mathcal{H}_{1}) =1$.
\end{prop}

\begin{IEEEproof}
	See \appref{appA}
\end{IEEEproof}

\begin{rmk} \label{rmk01}
	The implementation of \propref{prop01} requires the knowledge of $\sigma_{\omega_{m}}^{2}$.
	The authors aware that this parameter should be estimated for a practical propagation environment.
	For instance, the estimation of Rician $K$-factor is an open research problem, where its estimation accuracy depends on system bandwidth.
	In scenarios with moving sensing targets, it is further affected by Doppler effects.
	Nevertheless, imperfect knowledge of $\sigma_{\omega_{m}}^{2}$ does not fundamentally alter the optimal decision rule presented in \propref{prop01}.
	Specifically, the estimated $\sigma_{\omega_{m}}^{2}$ can be directly substituted into \eqref{eqn11031105}.
	Due to page limitation, the investigation of $\sigma_{\omega_{m}}^{2}$ estimation techniques will be presented in the forthcoming journal version.
\end{rmk}

\begin{rmk} \label{rmk02}
	In \propref{prop01}, the prior LoS probability is essential for optimal decision-making.
	However, there exists a fundamental challenge from the circular dependency: determining the environmental layout requires knowledge of prior probabilities, while these probabilities depend on the environment layout itself.
	In sensing applications, this challenge is typically addressed by replacing $\Theta_{m}$ with an alternative threshold determined by the desired false alarm probability.
	In practice, a common approach involves setting $\mathcal{P}(\mathcal{H}_{0}) = \mathcal{P}(\mathcal{H}_{1})$ \cite{Cook2012}.
	Based on this assumption, the decision rule and error rate of LoS state estimation are presented in \corref{cor01} and \corref{cor02}, respectively.
\end{rmk}

\begin{figure*}[t]
	\begin{minipage}[t]{1\textwidth}
		\centering
		\hspace{0.2em}
		\includegraphics[width=0.55\textwidth]{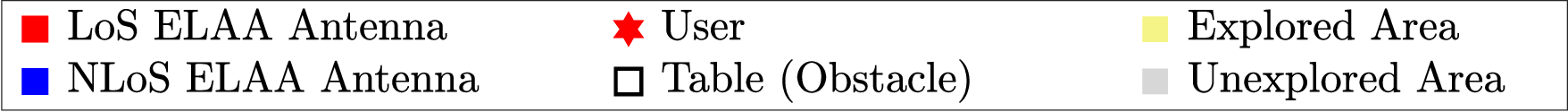}			   	
	\end{minipage}
	\subfigure[Initialization Map (No \gls{mt})]{
		\begin{minipage}[t]{0.239\textwidth}
			\label{fig02a}
			\centering
			\vspace{-0.5em}		
			\includegraphics[width=0.99\textwidth]{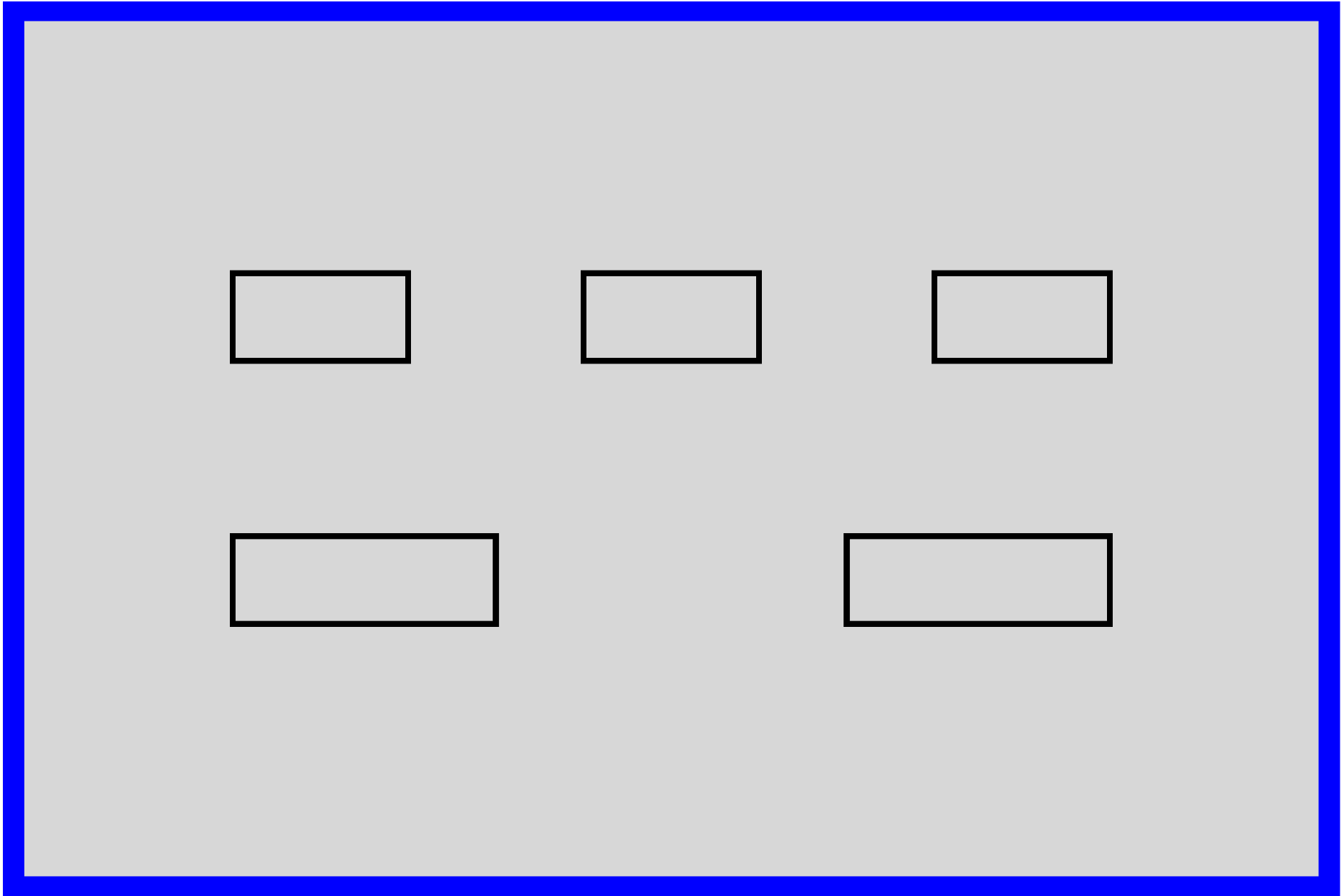}
			\vspace{-2.2em}				   	
	\end{minipage}}		
	\subfigure[\Gls{mt} location 1]{
		\begin{minipage}[t]{0.239\textwidth}
			\label{fig02b}
			\centering
			\vspace{-0.5em}		
			\includegraphics[width=0.99\textwidth]{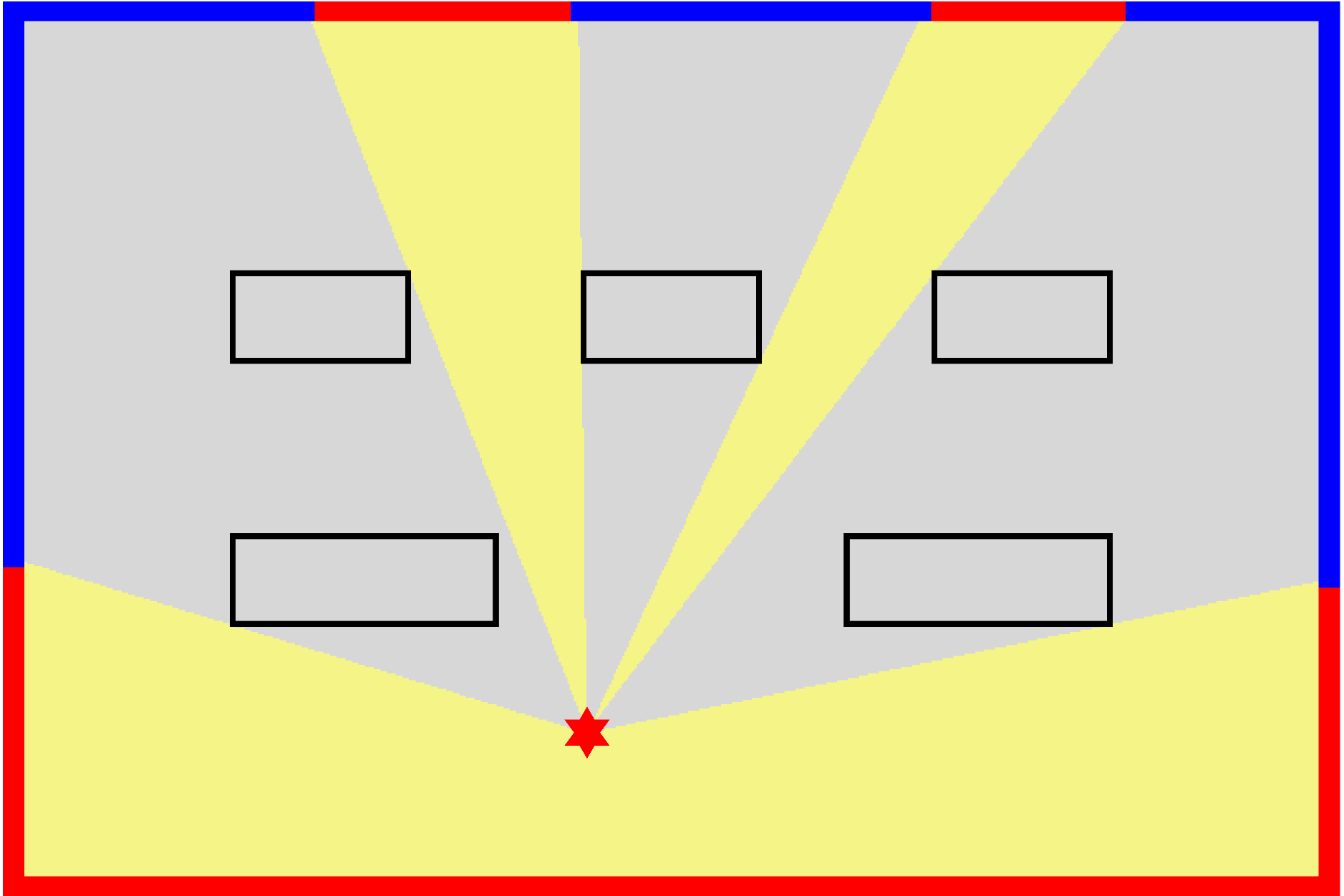}
			\vspace{-2.2em}				
	\end{minipage}}
	\subfigure[\Gls{mt} location 2]{
		\begin{minipage}[t]{0.239\textwidth}
			\label{fig02c}
			\centering
			\vspace{-0.5em}		
			\includegraphics[width=0.99\textwidth]{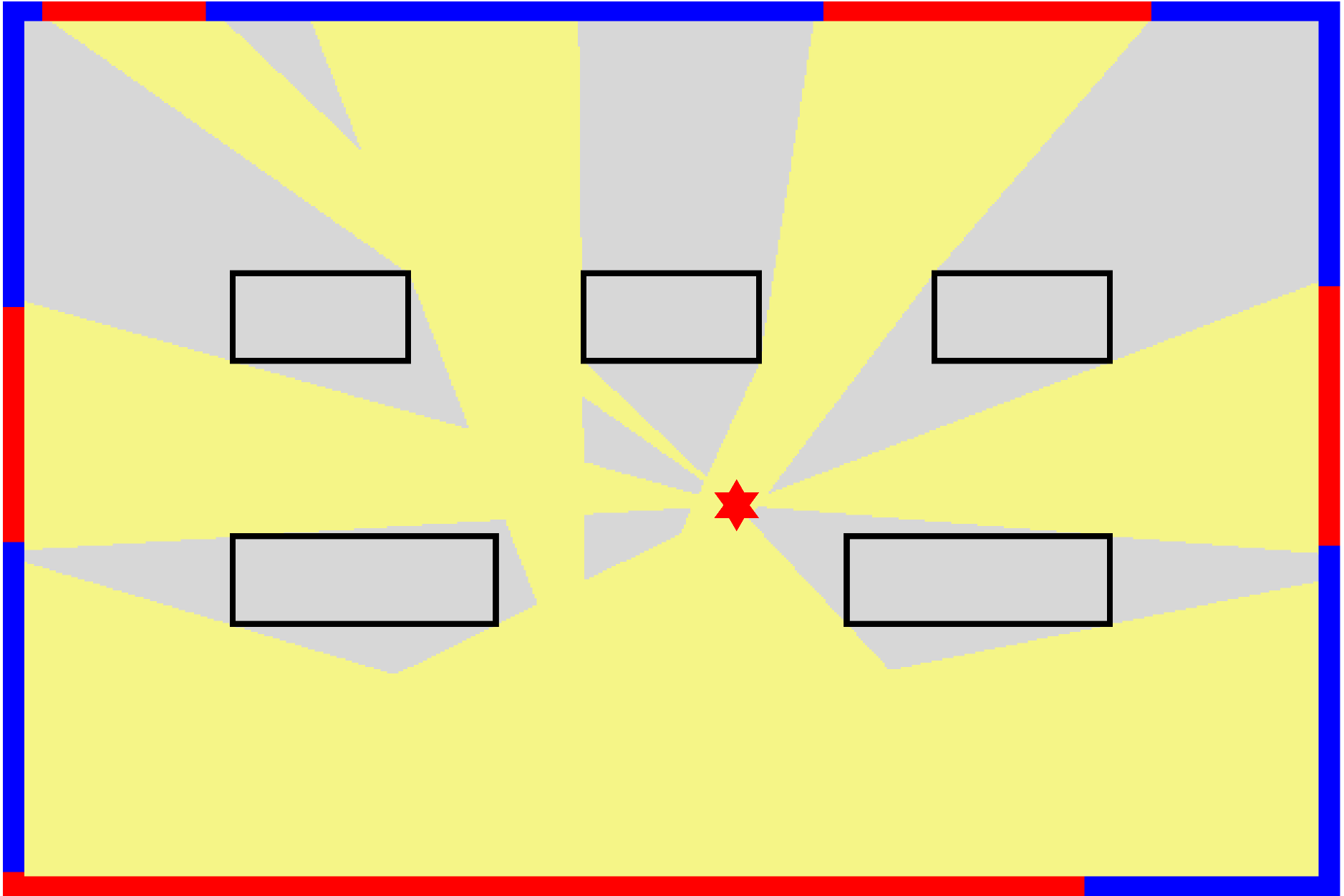}
			\vspace{-2.2em}				   	
	\end{minipage}}
	\subfigure[\Gls{mt} location 3]{
		\begin{minipage}[t]{0.239\textwidth}
			\label{fig02d}
			\centering
			\vspace{-0.5em}		
			\includegraphics[width=0.99\textwidth]{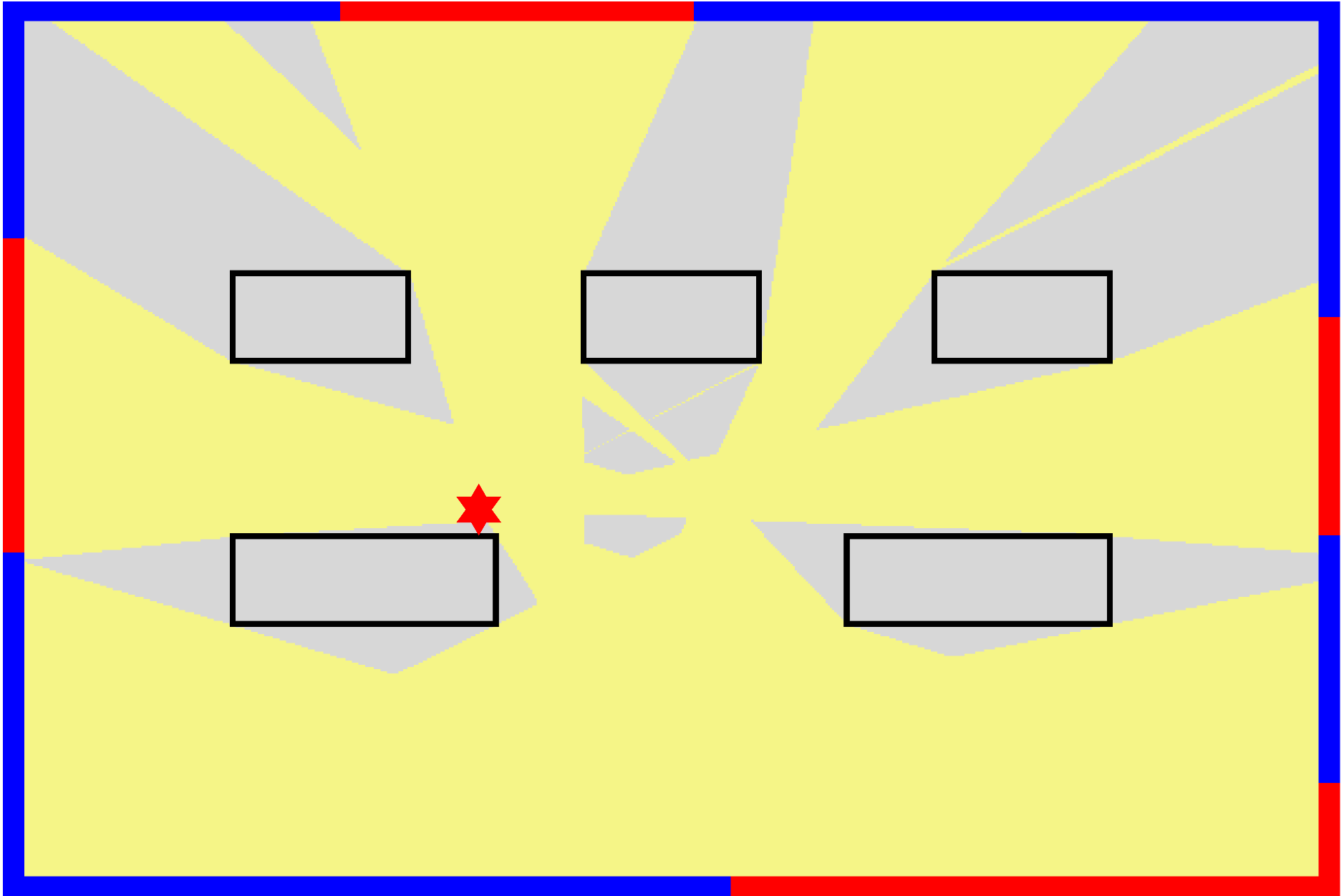}
			\vspace{-2.2em}				   	
	\end{minipage}}
	\subfigure[\Gls{mt} location 4]{
		\begin{minipage}[t]{0.239\textwidth}
			\label{fig02e}
			\centering
			\includegraphics[width=0.99\textwidth]{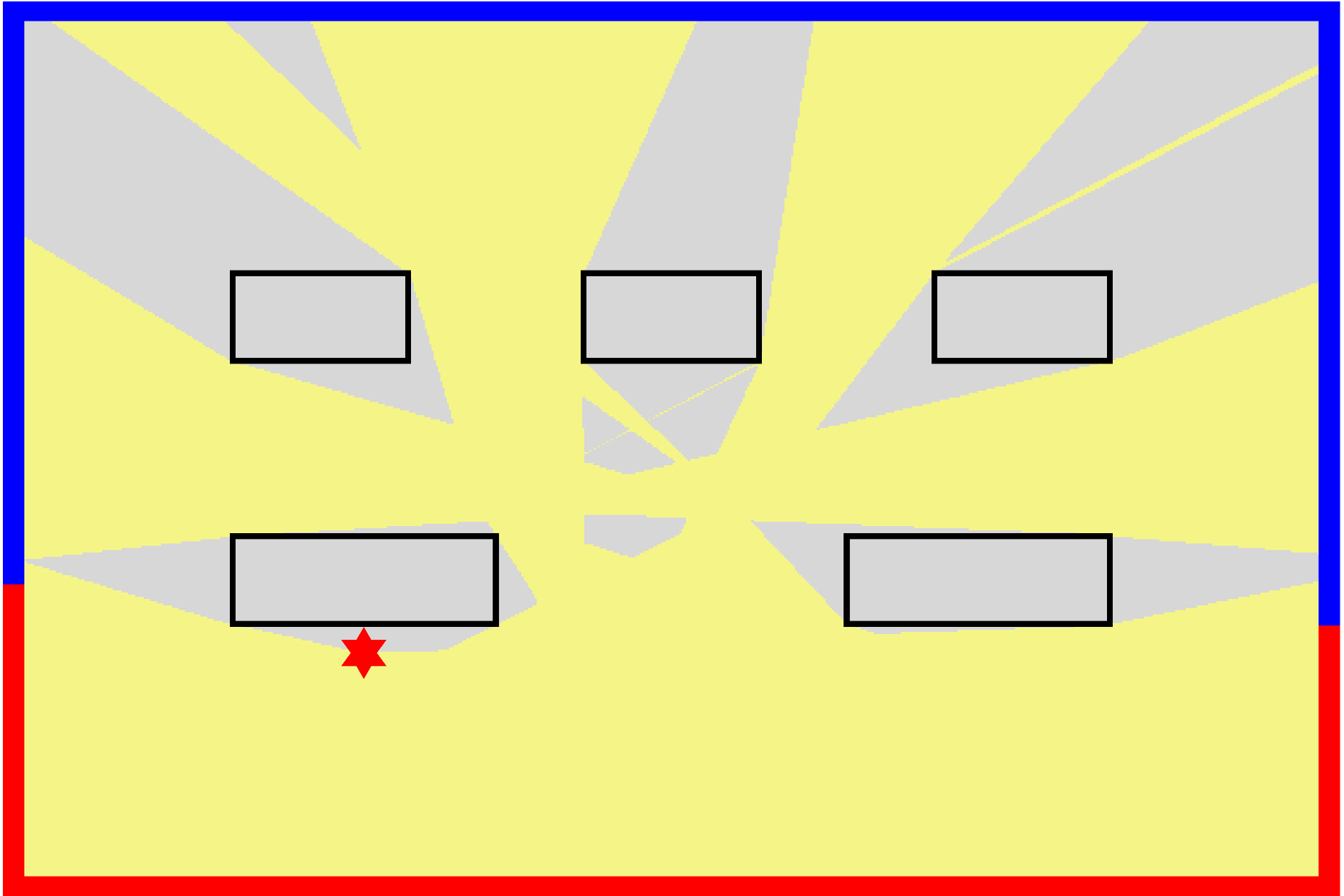}
			\vspace{-2.2em}				   	
	\end{minipage}}
	\subfigure[\Gls{mt} location 5]{
		\begin{minipage}[t]{0.239\textwidth}
			\label{fig02f}
			\centering
			\includegraphics[width=0.99\textwidth]{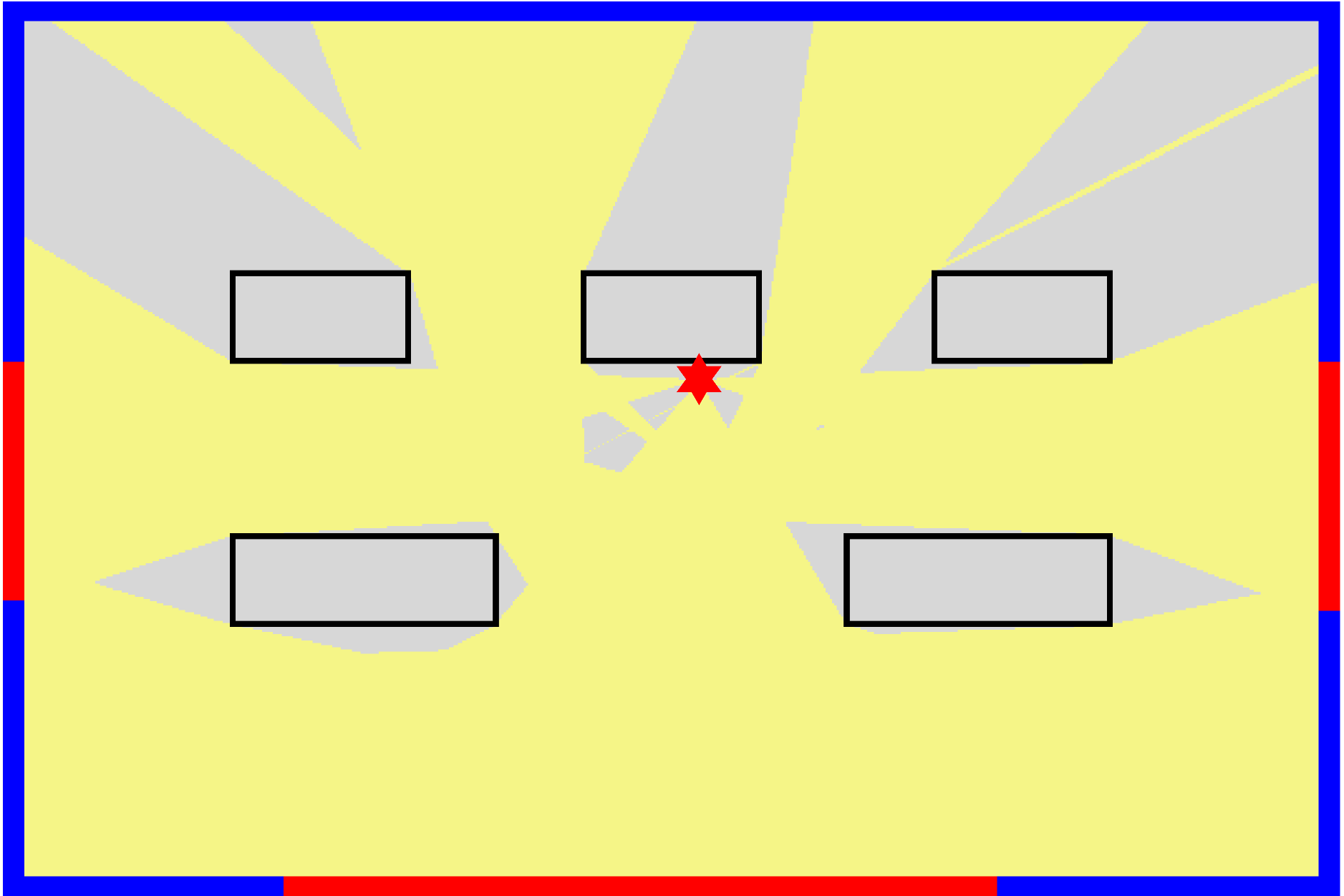}
			\vspace{-2.2em}				
	\end{minipage}}
	\subfigure[\Gls{mt} location 6]{
		\begin{minipage}[t]{0.239\textwidth}
			\label{fig02g}
			\centering
			\includegraphics[width=0.99\textwidth]{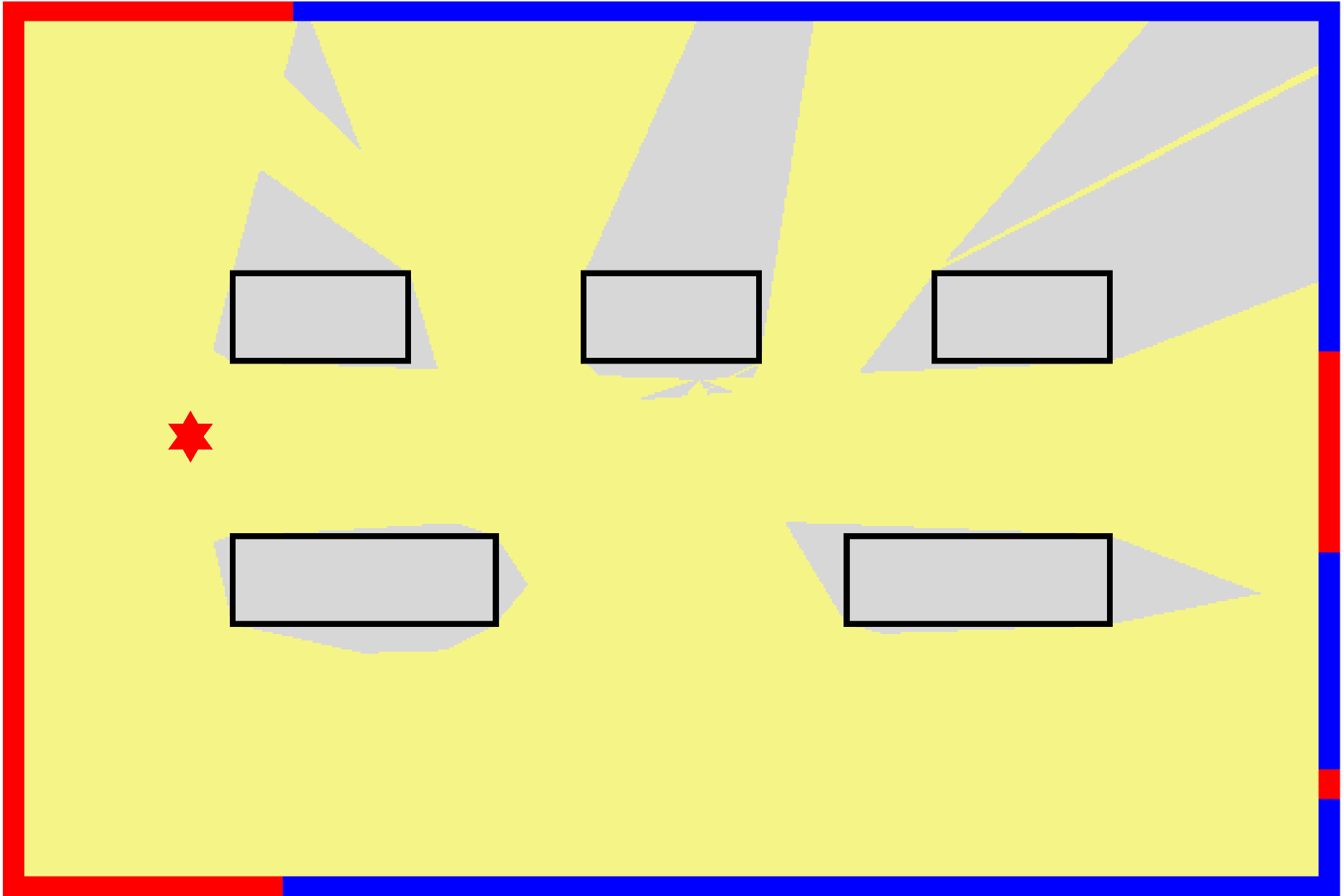}
			\vspace{-2.2em}				   	
	\end{minipage}}
	\subfigure[\Gls{mt} location 7]{
		\begin{minipage}[t]{0.239\textwidth}
			\label{fig02h}
			\centering
			\includegraphics[width=0.99\textwidth]{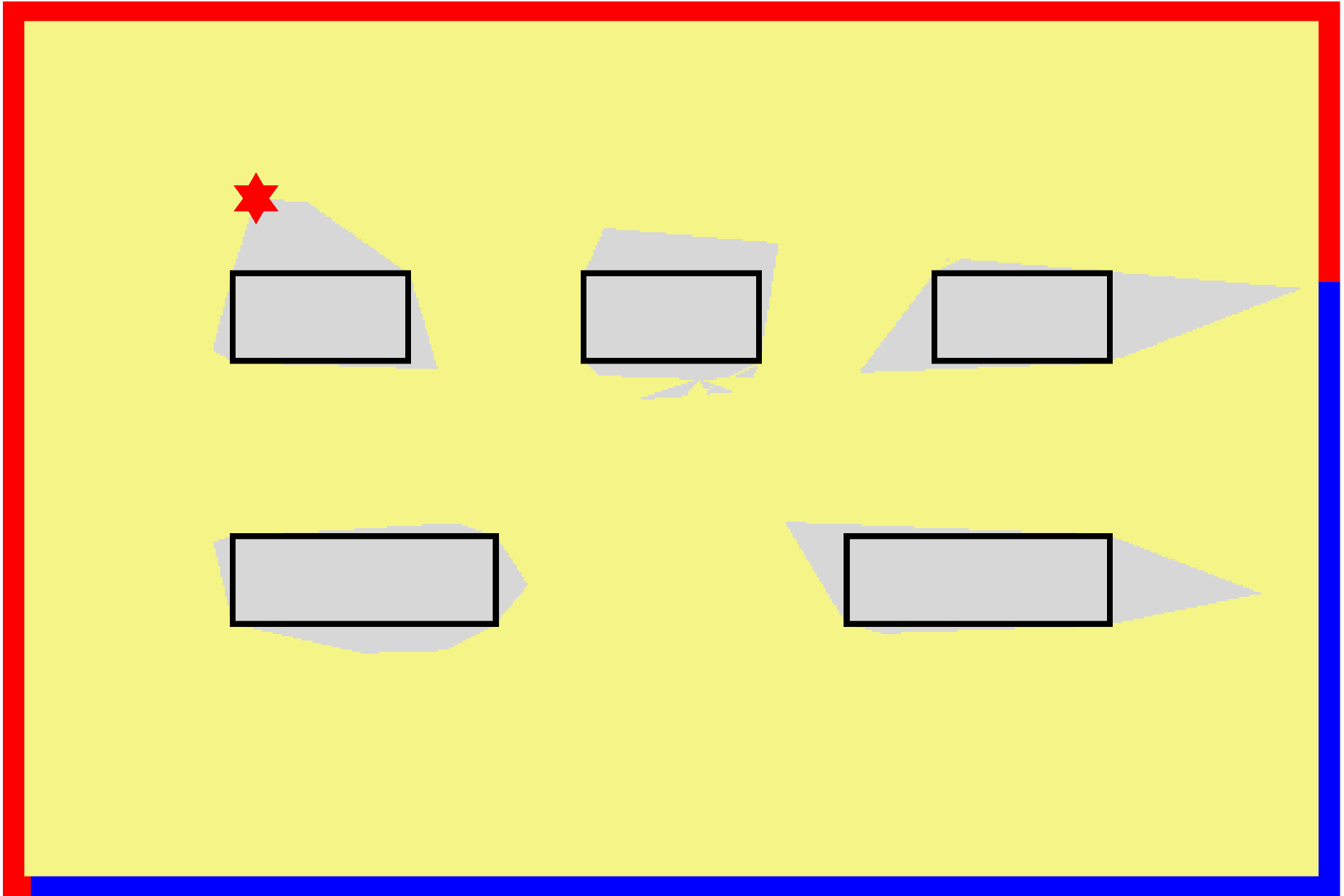}
			\vspace{-2.2em}				   	
	\end{minipage}}
	\caption{\label{fig02} Progressive environmental mapping with perfect \gls{los} information across $7$ \gls{mt} locations. The mapped area incrementally approximates the actual environment as the user traverses locations 1-7, illustrating enhanced spatial perception through increased sampling points.}
	\vspace{-1em}
\end{figure*}

\begin{cor} \label{cor01}
	Given $\hat{h}_{m}$, assuming $\mathcal{P}(\mathcal{H}_{0}) = \mathcal{P}(\mathcal{H}_{1})$, the optimal decision rule in \propref{prop01} is simplified to
	\begin{equation} \label{eqn05330925}
		\hat{b}_{m} =\Big(\big|\hat{h}_{m}\big|^{2} - \big|\hat{h}_{m} - 	\phi_{m}\big|^{2}\Big) \underset{\mathcal{H}_{0}}{\overset{\mathcal{H}_{1}}{\gtrless}} 0,
	\end{equation}
\end{cor}

\begin{IEEEproof}
 	The proof follows directly by substituting $\mathcal{P}(\mathcal{H}_{0}) = \mathcal{P}(\mathcal{H}_{1})$ into \eqref{eqn11031105}.
\end{IEEEproof}

\begin{rmk}
\corref{cor01} presents a remarkably simplified yet powerful decision rule for LoS state estimation.
This rule reduces to comparing the difference between $|\hat{h}_{m} - \phi_{m}|^{2}$ and $|\hat{h}_{m}|^{2}$ against a threshold of zero.
A positive difference indicates $\mathcal{H}_{1}$, suggesting a LoS component, while a non-positive difference signifies $\mathcal{H}_{0}$, implying an NLoS condition.
As discussed in \remref{rmk01}, $\sigma_{\omega_{m}}^{2}$ is often challenging to obtain accurately in practical scenarios.
Hence, a key advantage of this decision rule lies in its independence from $\sigma_{\omega_{m}}^{2}$.
\end{rmk}

\begin{cor} \label{cor02}
For a given \gls{mt} location, the error probability of LoS state estimation using the decision rule in \corref{cor01} can be expressed as follows
	\begin{equation} \label{eqn06040925}
		\mathscr{E} = \dfrac{1}{M} \sum_{m = 1}^{M} \mathcal{Q} \Bigg(\sqrt{\frac{Kp_{m}}{2p_{m} + 2(K + 1)\sigma_{v}^{2}}}\ \Bigg),
	\end{equation}
	where $\mathcal{Q}(\cdot)$ represents the Gaussian Q-function.
\end{cor}

\begin{IEEEproof}
	See \appref{appB}
\end{IEEEproof}

\begin{rmk} \label{rmk04}
	\corref{cor02} indicates that $\mathscr{E}$ is determined by three critical factors: the Rician $K$-factor, path-loss, and channel estimation error.
	In practical communication scenarios, the utilization of longer pilot sequences can effectively reduce channel estimation errors, i.e., $\sigma_{v}^{2}$.
	Hence, considering an ideal case where $\sigma_{v}^{2} = 0$, $\mathscr{E}$ simplifies to $\mathcal{Q}(\sqrt{K/2})$. 
	This demonstrates a direct relationship between error probability and the Rician $K$-factor, independent of the path-loss.
	In high-frequency bands such as mmWave and THz, the LoS component typically dominates the wireless channel \cite{Ge2023a}.
	Moreover, for a given carrier frequency, a higher $K$-factor can be obtained in the first channel tap by utilizing higher bandwidth.
	Such discussion will be presented in the journal paper due to spacing constraints.
\end{rmk}

Next, the proposed environmental mapping method utilizing the LoS state of ELAA channel is presented.

\subsection{Environmental Mapping Algorithm}
In this section, the proposed environmental mapping method is visualized step-by-step.
The fundamental principle of this method is based on the fact that \gls{los} links between the \gls{mt} and \gls{elaa} antennas indicate obstacle-free paths.

{\it Environment layout:}
A detailed illustration of the environment layout targeted for mapping is presented in \figref{fig02a}.
The setting is an indoor office space, precisely measured at $15$ meters in width and $25$ meters in length.
A rectangular \gls{elaa} is installed with its antennas uniformly distributed along the four walls of the office.
In this example, there are $M = 256$ service antennas.
The office interior includes potential obstacles, represented by five rectangular tables.

The proposed method aims to comprehensively explore and precisely map the office layout. \figref{fig02} illustrates the progressive mapping steps of this method.
Sub-figure (a) depicts the initialization step, showing an unexplored environment with five rectangular obstacles representing tables.
Sub-figures (b-h) demonstrate the mapping process as a \gls{mt} traverses seven different locations within the room.
At each \gls{mt} location, the entire map can be divided into $M$ triangular areas.
Colored squares along the perimeter of each sub-figure represent the \gls{elaa} antennas, with red indicating \gls{los} and blue denoting \gls{nlos} conditions relative to the \gls{mt}'s current position.
The algorithm updates the environment map based on \gls{los} information as the \gls{mt} moves.
It delineates the triangular areas to be explored when the link between the \gls{mt} and corresponding \gls{elaa} antenna is in \gls{los} state. 
The progression from sub-figures (b) to (h) exhibits a gradual increase in the explored area, as the algorithm infers obstacle presence along paths with \gls{nlos} conditions.
Sub-figure (h) displays a nearly complete environmental map, closely resembling the actual layout shown in the initialization map.

This visualization demonstrates how the method utilizes ELAA spatial resolution and MT mobility to accurately outline the environment, with perfect LoS state.
It is crucial to note, however, that the proposed method functions effectively with both perfect and imperfect \gls{los} state information.
The subsequent section will demonstrate the impact of imperfect \gls{los} state information on the mapping performance.

\section{Simulation and Numerical Results} \label{sec05}
In this section, the objectives are to:
\textit{1)} examine how the number of ELAA antennas and \gls{mt} locations influence mapping accuracy;
\textit{2)} validate the theoretical error probability of LoS state estimation; 
\textit{3)} investigate the impact of LoS estimation error on mapping performance.
The accuracy metric is chosen as \gls{iou}, a measure commonly employed in target detection tasks \cite{Rezatofighi2019}.
Mathematically, it is expressed as follows
\begin{equation}
	\mathrm{IoU} = \dfrac{A_{\text{unexp}} \cap A_{\text{table}}}{A_{\text{unexp}} \cup A_{\text{table}}} \times 100 \%,
\end{equation}
where $A_{\text{unexp}}$ indicates the unexplored area (i.e., gray color in \figref{fig02}) potentially containing obstacles, and $A_{\text{table}}$ represents the actual area occupied by tables.
The environment layout employed for all experiments corresponds to that depicted in \figref{fig02}.
The \gls{mt} locations are assumed to follow uniform distribution in the environment.
Each experiment undergoes sufficient repetitions to compute the average-\gls{iou}.
These three objectives establish the following three case studies.

\begin{figure}[t]
	\subfigure{
		\begin{minipage}[t]{0.49\textwidth}
			\centering
			\includegraphics[width=0.90\textwidth]{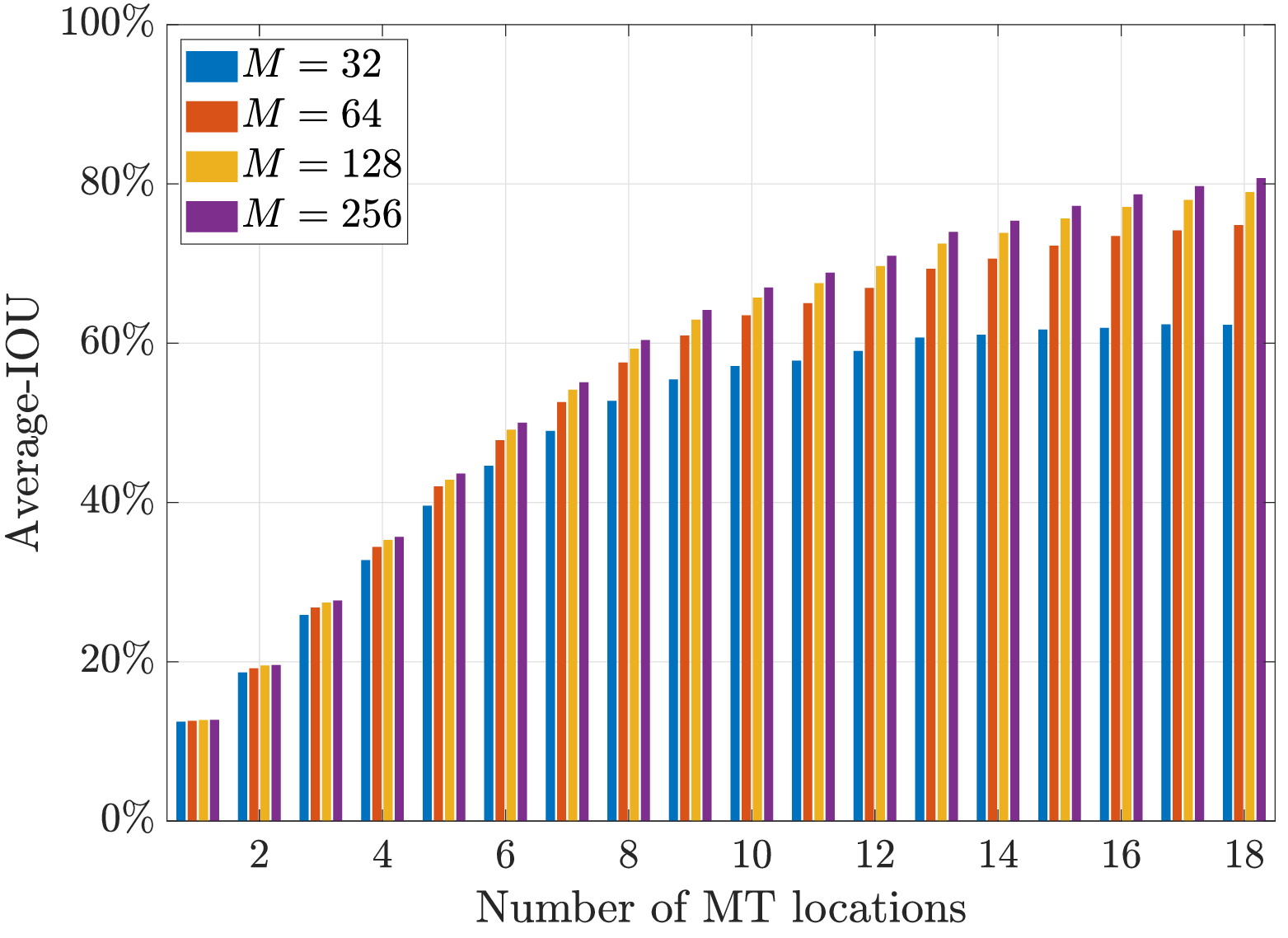}  	
	\end{minipage}}
	\caption{\label{fig03} Average-IoU performance versus the number of \gls{mt} locations for varying quantities of ELAA antennas. It can be found that increasing the number of antennas enhances environmental mapping accuracy.}	
	\vspace{-0em}
\end{figure}

{\it Case Study 1:}
The objective of this case study is to investigate the relationship between the \gls{iou} metric and the number of \gls{elaa} antennas, assuming perfect \gls{los} state information. 
\figref{fig03} presents the average-IoU performance as a function of the number of \gls{mt} locations for various antenna configurations ($M = 32$, $64$, $128$, and $256$). 
The results demonstrate that the IoU improves and gradually converges to a saturation point as the number of observation locations increases.
For instance, with $18$ \gls{mt} locations, the IoU increases from approximately $60\%$ for $32$ antennas to $80\%$ for $256$ antennas.
Moreover, the IoU performance exhibits significant improvement with an increasing number of ELAA antennas.
This improvement can be attributed to the direct relationship between the number of antennas and the spatial resolution provided by the ELAA.
In short, denser arrays allow for finer spatial sampling of the environment, resulting in more accurate maps.

\begin{figure}[t]
	\centering
	\subfigure{
		\begin{minipage}[t]{0.49\textwidth}
			\centering
			\includegraphics[width=0.90\textwidth]{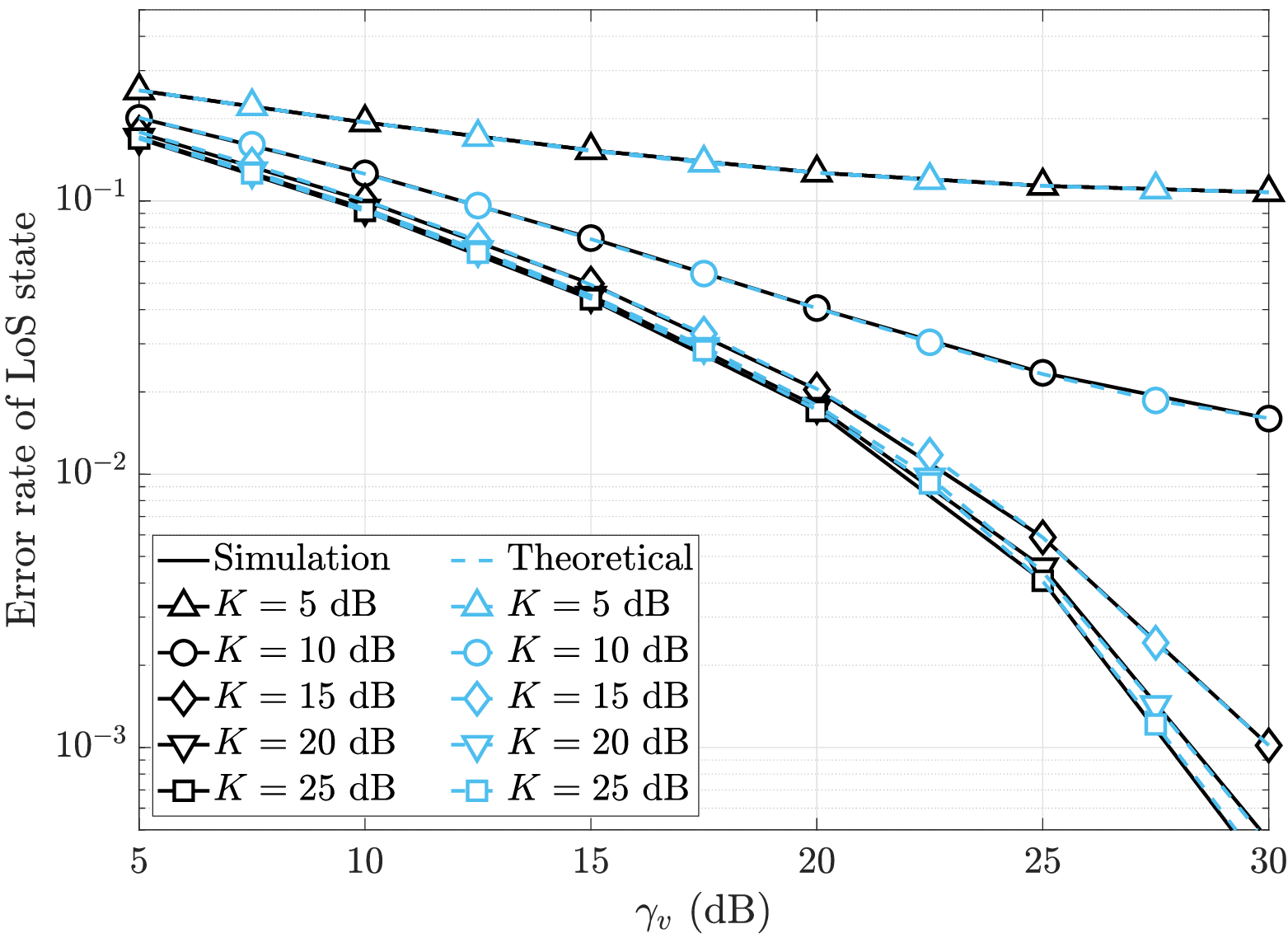} 	
	\end{minipage}}
	\caption{\label{fig04} Error rate of LoS state estimation as a function of $\gamma_{v}$ for different Rician $K$-factors. The simulation results closely match the theoretical error probability derived in \corref{cor02}.}
	\vspace{-1em}
\end{figure}

{\it Case Study 2:}
This case study aims to validate the theoretical error probability of LoS state estimation using simulation results.
Inaccuracies in LoS estimation arise from channel estimation errors and random \gls{nlos} components.
The ratio of NLoS component is measured by the Rician $K$-factor.
To measure the channel estimation error, the following parameter for each MT location is defined
\begin{equation}
	\gamma_{v} \triangleq \dfrac{\mathbb{E}\{\|\mathbf{h}\|^{2}\}}{M\sigma_{v}^{2}},
\end{equation}
where $\mathbb{E}\{\cdot\}$ represents the expectation of the input value.
A higher $\gamma_{v}$ indicates smaller channel estimation error.

In \figref{fig04}, the error probability of LoS state estimation as a function of $\gamma_{v}$ for various $K$ values is illustrated.
As anticipated, the error probability decreases as $\gamma_{v}$ increases across all considered $K$-factor values.
Higher $K$ result in lower error probabilities, indicating that stronger LoS components enhance the reliability of LoS state estimation.
The simulation results closely align with the theoretical curves, validating the accuracy of the theoretical analysis.
This indicates that the derived error probability expressions provide a reliable method for predicting LoS state estimation performance in practical scenarios.
Notably, even at high $\gamma_{v}$ values (e.g., $30$ dB), the error probability remains non-negligible, particularly for lower $K$-factor values.
This observation highlights the inherent challenge in accurately determining the LoS state in the presence of NLoS components.
For the narrowband case, implementing the proposed method in higher frequency bands, e.g., mmWave and THz, is recommended, as the LoS path typically dominates in these wireless channels.
As discussed in \remref{rmk01} and \remref{rmk04}, future research may explore wideband signals as a potential strategy to further reduce the influence of NLoS components.

\begin{figure*}[t]
	\centering
	\subfigure[Number of \gls{mt} locations: $8$]{
		\begin{minipage}[t]{0.49\textwidth}
			\label{fig05a}
			\centering
			\includegraphics[width=0.90\textwidth]{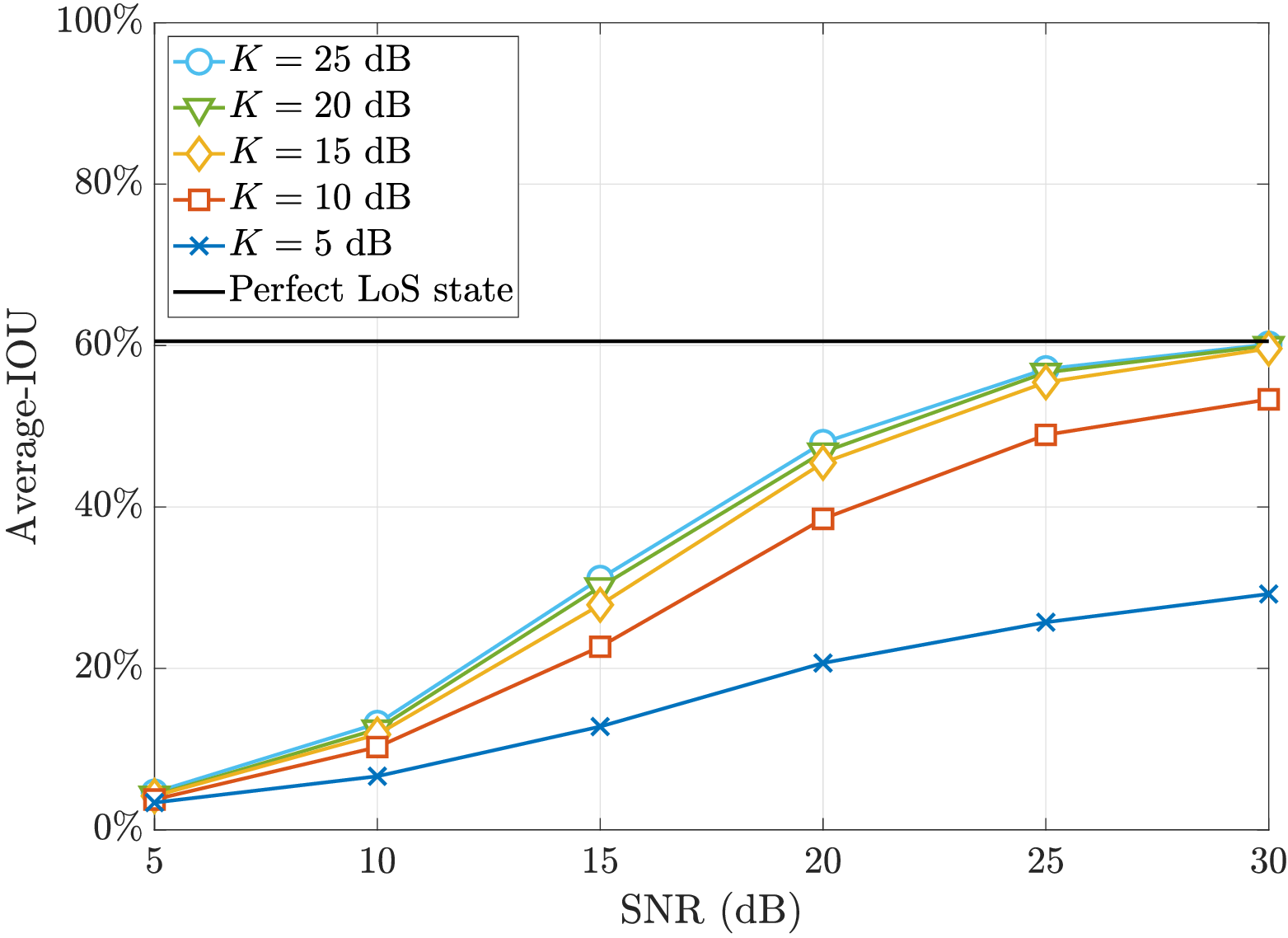}
			\vspace{0.5em}				   	
	\end{minipage}}
	\subfigure[Number of \gls{mt} locations: $18$]{
		\begin{minipage}[t]{0.49\textwidth}
			\label{fig05b}
			\centering
			\includegraphics[width=0.90\textwidth]{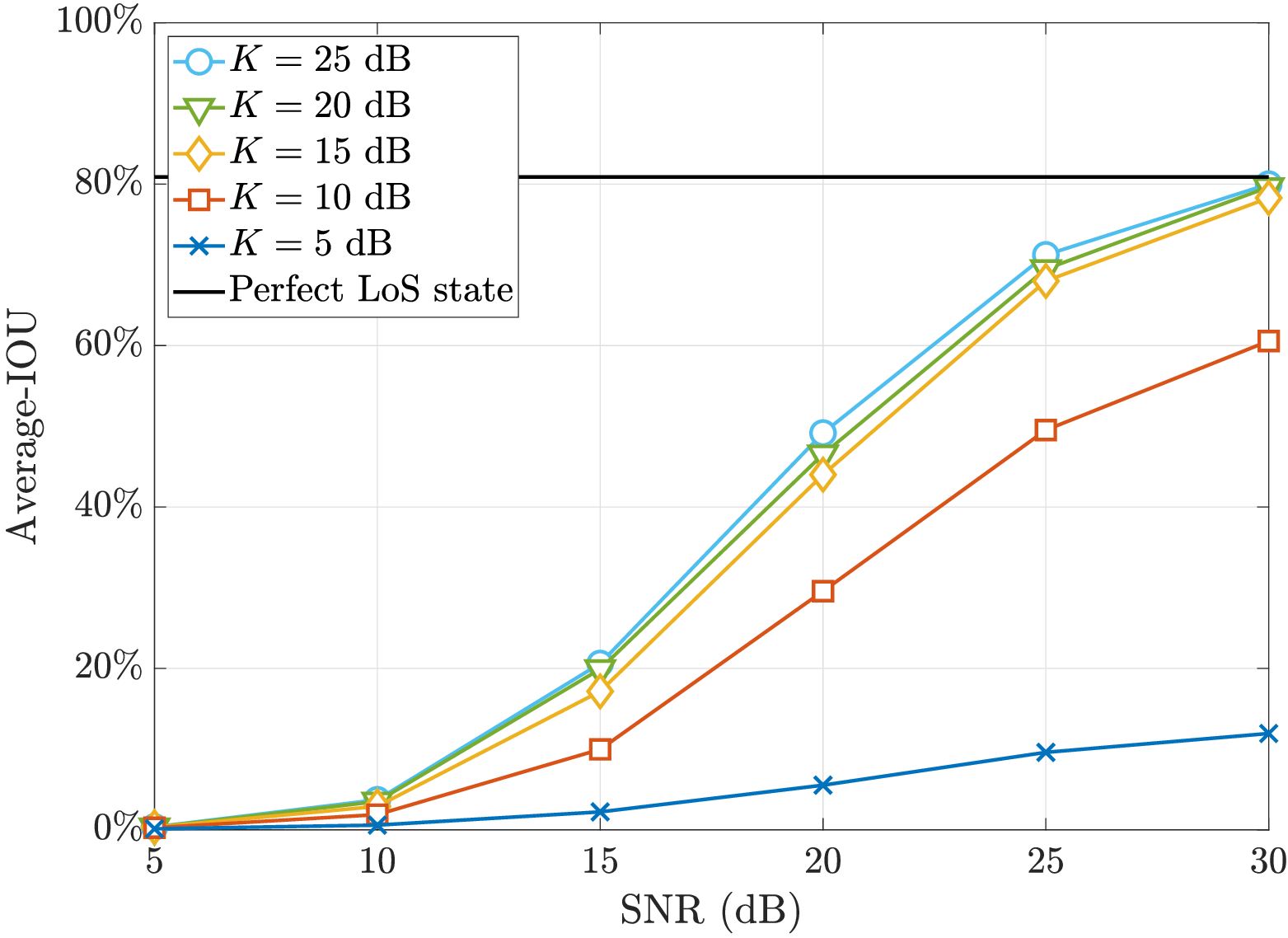}
			\vspace{0.5em}				
	\end{minipage}}
	\caption{\label{fig05} Average-IoU performance as a function of $\gamma_{v}$ for different values of the Rician $K$-factor. The results demonstrate an inverse relationship between the mapping accuracy and the magnitude of NLoS components plus channel estimation errors.}
	\vspace{-0.5em}
\end{figure*}

{\it Case Study 3:}
This case study aims to examine the effect of \gls{los} estimation error on environmental mapping accuracy.
In \figref{fig05}, the number of ELAA antennas is set to be $256$ as an example.
The figure is composed of two sub-figures, depicting scenarios with $8$ and $16$ \gls{mt} locations respectively.
Each sub-figure displays the average-IoU versus $\gamma_{v}$ for $K$ ranging from $5$ dB to $25$ dB.
Both sub-figures demonstrate that average-IoU rises with increasing $\gamma_{v}$ values, indicating enhanced mapping performance as signal quality improves.
A solid line represents the perfect LoS state, setting an upper bound of mapping performance.
Comparing the two sub-figures reveals that increasing the number of \gls{mt} locations from $8$ to $18$ results in higher average-\gls{iou}.
These results align with the findings presented in \figref{fig03} of \textit{Case Study 1}.
Moreover, comparison between \figref{fig04} (in \textit{Case Study 2}) and \figref{fig05} reveals a direct correlation between mapping accuracy and the error probability of LoS state estimation.
Investigation of the theoretical relationship between IoU and error probability is an interesting direction for future research.

In \figref{fig05}, higher $K$ values consistently produce better performance across all $\gamma_{v}$ values.
This suggests that stronger LoS components contribute to more precise mapping.
Specifically, for the case $K = 25$ dB and $\gamma_{v} = 30$ dB, the mapping accuracy is close to that with perfect LoS state information.
The accuracy gap between different $K$ values becomes more pronounced at lower channel estimation errors.
As discussed before, channel estimation error can be reduced by sending pilot sequences with more power.
This indicates that the Rician $K$-factor represents the primary limiting factor in the narrowband scenarios.
Therefore, implementation of this algorithm is particularly effective in LoS-dominant environments, such as those utilizing higher carrier frequency.
Notably, employing wideband signals for channel estimation can isolate the direct LoS path more effectively, enhancing its dominance.

\section{Conclusion and Outlook}
This paper presents a novel environmental mapping method utilizing the \gls{los} state information of \gls{elaa} channels.
The proposed method is a communication-centric \gls{isac} paradigm, enabling environmental mapping capabilities while maintaining uncompromised communication performance.
\gls{los} state estimation is formulated as a binary hypothesis testing problem, with both optimal and practical solutions derived.
A mathematical analysis of the theoretical error probability for \gls{los} state estimation is provided, showing close alignment with simulation results.
Moreover, the proposed mapping method leverages ELAA spatial resolution and \gls{mt} mobility to infer obstacle presence and location.
Our simulation results demonstrate that the approach can efficiently outline the layout of the environment, achieving a high \gls{iou} exceeding $80\%$ when utilizing $256$ service antennas and $18$ MT locations.

Future work will focus on enhancing LoS state estimation accuracy through wideband signal utilization.
Additional research directions include analyzing system performance with moving objects, optimizing user movement trajectories, addressing imperfect MT localization, and extending the framework to \gls{3d} environments.
These investigations aim to advance the practical deployment and applicability of this approach in real-world settings.

\appendices
\section{Proof of \propref{prop01}} \label{appA}
To LoS state estimation problem can be solved by the following binary hypothesis testing \cite{Cook2012}
\begin{equation} \label{eqn05120925}
	\hat{b}_{m} = \frac{\mathcal{P}\big(\hat{h}_{m}|\mathcal{H}_{1}\big)}{\mathcal{P}\big(\hat{h}_{m}|\mathcal{H}_{0}\big)} \underset{\mathcal{H}_{0}}{\overset{\mathcal{H}_{1}}{\gtrless}} \frac{\mathcal{P}\big(\mathcal{H}_{0}\big)}{\mathcal{P}\big(\mathcal{H}_{1}\big)}.
\end{equation}
where $\mathcal{P}(\hat{h}_{m}|\mathcal{H}_{0})$ and $\mathcal{P}(\hat{h}_{m}|\mathcal{H}_{1})$ denote the \gls{pdf} of $\hat{h}_{m}$ condition on $\mathcal{H}_{0}$ and $\mathcal{H}_{1}$, respectively.
Under hypothesis $\mathcal{H}_{0}$, \eqref{eqn09270915} becomes $\hat{h}_{m} = \omega_{m}$.
Therefore, we have the following
\begin{equation}\label{eqn01150925}
	\mathcal{P}\big(\hat{h}_{m}|\mathcal{H}_{0}\big) = \frac{1}{\pi\sigma_{\omega_{m}}^{2}}
	\exp\bigg(-\dfrac{|\hat{h}_{m}|^{2}}{\sigma_{\omega_{m}}^{2}}\bigg).
\end{equation}
Under hypothesis $\mathcal{H}_{1}$, \eqref{eqn09270915} becomes $\hat{h}_m = \phi_m + \omega_m$, and $\mathcal{P}(\hat{h}_{m}|\mathcal{H}_{1})$ can be expressed as follows
\begin{equation}\label{eqn01160925}
	\mathcal{P}\big(\hat{h}_{m}|\mathcal{H}_{1}\big) = \frac{1}{\pi\sigma_{\omega_{m}}^{2}}
	\exp\bigg(-\dfrac{|\hat{h}_{m} - \phi_{m}|^{2}}{\sigma_{\omega_{m}}^{2}}\bigg).
\end{equation}
Substituting \eqref{eqn01150925} and \eqref{eqn01160925} into \eqref{eqn05120925},  taking the logarithm of both sides and rearranging the terms, the decision rule in \propref{prop01} is therefore obtained.

\section{Proof of \corref{cor02}} \label{appB}
The average error probability of estimating LoS state for a given \gls{mt} location can be expressed as follows
\begin{equation} \label{eqn09291010}
	\mathscr{E} = \dfrac{1}{M} \sum_{m = 1}^{M} \mathscr{E}_{m},
\end{equation}
where $\mathscr{E}_{m} \triangleq \mathcal{P}(\hat{b}_{m} \neq b_{m})$ represent the error probability of the $m^{th}$ link, and it can be expressed as follows
\begin{equation}\label{eqn03291105}
	\mathscr{E}_{m} = \mathcal{P}\big(\mathcal{H}_{0}\big)\mathcal{P}\big(\hat{b}_{m} = 1 | \mathcal{H}_{0}\big) + \mathcal{P}\big(\mathcal{H}_{1}\big)\mathcal{P}\big(\hat{b}_{m} = 0 | \mathcal{H}_{1}\big).
\end{equation}
According to \propref{prop01}, we have the following
\begin{subequations}\label{eqn02090926}
	\begin{align}
		\mathcal{P}\big(\hat{b}_{m} = 1 | \mathcal{H}_{0}\big) = \mathcal{P}\big(|\hat{h}_{m}|^{2} - |\hat{h}_{m} - \phi_{m}|^{2} > 0 | \mathcal{H}_{0}\big); \label{eqn02090926a} \\
		\mathcal{P}\big(\hat{b}_{m} = 0 | \mathcal{H}_{1}\big) = \mathcal{P}\big(|\hat{h}_{m}|^{2} - |\hat{h}_{m} - \phi_{m}|^{2} < 0 | \mathcal{H}_{1}\big). \label{eqn02090926b}
	\end{align}
\end{subequations}
Simplifying \eqref{eqn02090926} yields
\begin{equation}
	\left.\begin{array}{l}
	\mathcal{P}\big(\hat{b}_{m} = 1 | \mathcal{H}_{0}\big)\\
	\mathcal{P}\big(\hat{b}_{m} = 0 | \mathcal{H}_{1}\big)
	\end{array}\right\}
	=\mathcal{P}\bigg(\Re\big(\omega_{m}\phi_m^{*}\big) < \frac{|\phi_{m}|^{2}}{2}\bigg),
\end{equation}
where $\Re(\omega_{m}\phi_{m}^{*})$ is a real Gaussian random variable with zero mean and variance $\frac{|\phi_{m}|^{2}\sigma_{\omega_{m}}^{2}}{2}$.
Standardizing this random variable and applying the Gaussian Q-function, results in
\begin{equation} \label{eqn06080925}
	\left.\begin{array}{l}
	\mathcal{P}\big(\hat{b}_{m} = 1 | \mathcal{H}_{0}\big)\\
	\mathcal{P}\big(\hat{b}_{m} = 0 | \mathcal{H}_{1}\big)
	\end{array}\right\}
	= \mathcal{Q}\bigg(\sqrt{\frac{|\phi_m|^2}{2\sigma_{\omega_{m}}^{2}}}\ \bigg).
\end{equation}
Plugging \eqref{eqn06080925} and $\mathcal{P}(\mathcal{H}_{0}) + \mathcal{P}(\mathcal{H}_{1}) =1$ into \eqref{eqn03291105} yields
\begin{equation} \label{eqn10511106}
	\mathscr{E}_{m}=\mathcal{Q}\bigg(\sqrt{\frac{|\phi_m|^2}{2\sigma_{\omega_{m}}^{2}}}\ \bigg)
\end{equation}
According to definition of $\phi_{m}$ in \eqref{eqn08220926}, it can be obtained that
\begin{equation} \label{eqn08240926}
	|\phi_{m}|^{2} =\dfrac{Kp_{m}}{K + 1}.
\end{equation}
Plugging \eqref{eqn08240926} and \eqref{eqn08300925} into \eqref{eqn10511106}, then plugging \eqref{eqn10511106} into \eqref{eqn09291010} yields the error probability in \corref{cor02}.

\section*{Acknowledgement}
This work was partially funded by the 5G/6GIC, University of Surrey, and partially by the UK DSIT under the TUDOR (Towards Ubiquitous 3D Open Resilient Network) project.

\ifCLASSOPTIONcaptionsoff
\newpage
\fi

\balance
\bibliographystyle{IEEEtran}
\bibliography{../IEEEabrv,../thesis_list} 
\begin{thebibliography}{10}
\providecommand{\url}[1]{#1}
\csname url@samestyle\endcsname
\providecommand{\newblock}{\relax}
\providecommand{\bibinfo}[2]{#2}
\providecommand{\BIBentrySTDinterwordspacing}{\spaceskip=0pt\relax}
\providecommand{\BIBentryALTinterwordstretchfactor}{4}
\providecommand{\BIBentryALTinterwordspacing}{\spaceskip=\fontdimen2\font plus
\BIBentryALTinterwordstretchfactor\fontdimen3\font minus
  \fontdimen4\font\relax}
\providecommand{\BIBforeignlanguage}[2]{{%
\expandafter\ifx\csname l@#1\endcsname\relax
\typeout{** WARNING: IEEEtran.bst: No hyphenation pattern has been}%
\typeout{** loaded for the language `#1'. Using the pattern for}%
\typeout{** the default language instead.}%
\else
\language=\csname l@#1\endcsname
\fi
#2}}
\providecommand{\BIBdecl}{\relax}
\BIBdecl

\bibitem{Liu2022b}
F.~Liu, Y.~Cui, C.~Masouros, J.~Xu, T.~X. Han, Y.~C. Eldar, and S.~Buzzi,
  ``Integrated sensing and communications: Toward dual-functional wireless
  networks for {6G} and beyond,'' \emph{IEEE J. Sel. Areas Commun.}, vol.~40,
  no.~6, pp. 1728--1767, Jun. 2022.

\bibitem{Cui2023}
M.~Cui, Z.~Wu, Y.~Lu, X.~Wei, and L.~Dai, ``Near-field {MIMO} communications
  for {6G}: {Fundamentals}, challenges, potentials, and future directions,''
  \emph{{IEEE} Commun. Mag.}, vol.~61, no.~1, pp. 40--46, Jan. 2023.

\bibitem{Liu2021}
J.~Liu, Y.~Ma, J.~Wang, N.~Yi, R.~Tafazolli, S.~Xue, and F.~Wang, ``A
  non-stationary channel model with correlated {NLoS/LoS} states for
  {ELAA-mMIMO},'' in \emph{Proc. IEEE Global Commun. Conf. (GLOBECOM)}, 2021,
  pp. 1--6.

\bibitem{Lu2023}
Y.~Lu and L.~Dai, ``Near-field channel estimation in mixed {LoS/NLoS}
  environments for extremely large-scale {MIMO} systems,'' \emph{{IEEE} Trans.
  Commun.}, vol.~71, no.~6, pp. 3694--3707, Jun. 2023.

\bibitem{Liu2024}
J.~Liu, Y.~Ma, A.~Elzanaty, and R.~Tafazolli, ``Near-field fading channel
  modeling for {ELAAs}: From communication to {ISAC},''
  \emph{arXiv:2401.17014}, Jan. 2024.

\bibitem{Qing2024}
C.~Qing, Z.~Liu, W.~Hu, Y.~Zhang, X.~Cai, and P.~Du, ``{LoS} sensing-based
  channel estimation in {UAV}-assisted {OFDM} systems,'' \emph{IEEE Wireless
  Commun. Lett.}, vol.~13, no.~5, pp. 1320--1324, 2024.

\bibitem{GonzalezPrelcic2024}
N.~González-Prelcic, M.~F. Keskin, O.~Kaltiokallio, M.~Valkama, D.~Dardari,
  X.~Shen, Y.~Shen, M.~Bayraktar, and H.~Wymeersch, ``The integrated sensing
  and communication revolution for {6G}: Vision, techniques,and applications,''
  \emph{Proc. IEEE}, pp. 1--0, 2024.

\bibitem{An2023}
J.~An, H.~Li, D.~W.~K. Ng, and C.~Yuen, ``Fundamental detection probability vs.
  achievable rate tradeoff in integrated sensing and communication systems,''
  \emph{IEEE Trans. Wireless Commun.}, vol.~22, no.~12, pp. 9835--9853, Dec.
  2023.

\bibitem{Zhi2024}
K.~Zhi, C.~Pan, H.~Ren, K.~K. Chai, C.-X. Wang, R.~Schober, and X.~You,
  ``Performance analysis and low-complexity design for {XL-MIMO} with
  near-field spatial non-stationarities,'' \emph{IEEE J. Sel. Areas Commun.},
  vol.~42, no.~6, pp. 1656--1672, Jun. 2024.

\bibitem{Elzanaty2024}
A.~Elzanaty, J.~Liu, A.~Guerra, F.~Guidi, Y.~Ma, and R.~Tafazolli, ``Near and
  far field model mismatch: Implications on {6G} communications, localization,
  and sensing,'' \emph{IEEE Internet of Things Mag.}, vol.~7, no.~5, pp.
  120--126, Sep. 2024.

\bibitem{Liu2024a}
J.~Liu, Y.~Ma, and R.~Tafazolli, ``A spatially non-stationary fading channel
  model for simulation and (semi-) analytical study of {ELAA-MIMO},''
  \emph{{IEEE} Trans. Wireless Commun.}, vol.~23, no.~5, pp. 5203--5218, May
  2024.

\bibitem{Amiri2022}
A.~Amiri, S.~Rezaie, C.~N. Manchón, and E.~de~Carvalho, ``Distributed receiver
  processing for extra-large {MIMO} arrays: A message passing approach,''
  \emph{{IEEE} Trans. Wireless Commun.}, vol.~21, no.~4, pp. 2654--2667, Apr.
  2022.

\bibitem{Liu2024b}
J.~Liu, J.~Wang, Y.~Ma, and R.~Tafazolli, ``Accelerating iteratively linear
  detectors in multi-user {(ELAA-)MIMO} systems with {UW-SVD},'' \emph{{IEEE}
  Trans. Wireless Commun.}, vol.~23, no.~11, pp. 16\,711--16\,724, Nov. 2024.

\bibitem{3GPP2022}
3GPP, ``{Study on channel model for frequencies from 0.5 to 100 GHz},'' {3rd
  Generation Partnership Project (3GPP)}, Technical Report (TR) 38.901, Mar.
  2022, version 17.0.0.

\bibitem{Cook2012}
C.~Cook, \emph{Radar signals: An introduction to theory and application}.\hskip
  1em plus 0.5em minus 0.4em\relax Elsevier, 2012.

\bibitem{Ge2023a}
Y.~Ge, H.~Kim, L.~Svensson, H.~Wymeersch, and S.~Sun, ``Integrated monostatic
  and bistatic {mmWave} sensing,'' in \emph{Proc. IEEE Global commun. Conf.
  (GLOBECOM)}, 2023, pp. 3897--3903.

\bibitem{Rezatofighi2019}
H.~Rezatofighi, N.~Tsoi, J.~Gwak, A.~Sadeghian, I.~Reid, and S.~Savarese,
  ``Generalized intersection over union: A metric and a loss for bounding box
  regression,'' in \emph{IEEE/CVF Conf. Comput. Vision Pattern Recognit.
  (CVPR)}, 2019, pp. 658--666.

\end{thebibliography}
\end{document}